\documentclass[twocolumn,aps,superscriptaddress,prb]{revtex4-2} 
\usepackage{amsmath}
\usepackage{graphicx} 

\begin{document}
\title{Possible gapless helical edge states in hydrogenated graphene}

\author{Yong-Cheng Jiang}
\affiliation{Research Center for Materials Nanoarchitectonics (MANA), National Institute for Materials Science (NIMS), Tsukuba 305-0044, Japan}
\affiliation{Graduate School of Science and Technology, University of Tsukuba, Tsukuba 305-8571, Japan}
\author{Toshikaze Kariyado}
\affiliation{Research Center for Materials Nanoarchitectonics (MANA), National Institute for Materials Science (NIMS), Tsukuba 305-0044, Japan}
\author{Xiao Hu}
\email{HU.Xiao@nims.go.jp}
\affiliation{Research Center for Materials Nanoarchitectonics (MANA), National Institute for Materials Science (NIMS), Tsukuba 305-0044, Japan}
\affiliation{Graduate School of Science and Technology, University of Tsukuba, Tsukuba 305-8571, Japan}

\date{\today}

\begin{abstract}
Electronic band structures in hydrogenated graphene are theoretically investigated by means of first-principle calculations and an effective tight-binding model. It is shown that regularly designed hydrogenation to graphene gives rise to a large band gap about 1~eV.  
Remarkably, by changing the spatial pattern of the hydrogenation, topologically distinct states can be realized, where the topological nontriviality is detected by $C_2$ parity indices in bulk and confirmed by the existence of gapless edge/interface states as protected by the mirror and sublattice symmetries. 
The analysis of the wave functions reveals that the helical edge states in hydrogenated graphene with the appropriate design carry pseudospin currents that are reminiscent of the quantum spin Hall effect. Our work shows the potential of hydrogenated graphene in pseudospin-based device applications.
\end{abstract}

\maketitle

\section{Introduction}

Graphene, atomically thin honeycomb network of carbon atoms, is regarded as a promising material for next-generation devices because of its intriguing properties~\cite{CastroNeto2009}. Graphene is tough and stiff due to its covalent nature of the carbon-carbon bonding. Also, graphene has high electronic mobility, which may lead to low loss and quick response electronic manipulation. On the other hand, the gapless band structure of graphene at the Fermi energy hinders its applications in conventional semiconductor devices, which requires a gap for electron manipulation~\cite{PhysRevLett.100.136804}. Approaches for inducing a gap in the band structure of graphene have been discussed for a long time~\cite{Son2006b,Han2007,Castro2007,Ponomarenko2008,Ni2008}. One possible way is to introduce a superstructure with an appropriate period. For instance, it is known that an adequately designed graphene nanomesh, i.e., graphene with regular array of holes~\cite{PhysRevLett.71.4389,PhysRevLett.100.136804,Bai2010,PhysRevB.98.195416}, acquires a band gap at the Fermi energy.

Importantly, pristine graphene is not just gapless, but its gapless band structure is described by a relativistic Dirac equation. This allows us to make gapped graphene topologically nontrivial. Topological phases of matter~\cite{Hasan2010,Qi2011,Weng2015,Bansil2016} have been one of the central topics in condensed matter physics and materials science. Topological states are often characterized by robust edge/interface modes~\cite{Hatsugai1993}. Because of the robustness, topological edge/interface modes potentially lead to interesting device applications. Theoretically, what is necessary for realizing topologically nontrivial states standing on the Dirac equation is a tunable mass term, especially, the sign change of the mass term. This is often declared as band inversion~\cite{Bernevig2006}. Namely, the sign flip of the mass term corresponds to an exchange of states at band edges clipping a focused band gap. In practice, the states are identified by their wave function symmetry, meaning that the band inversion is detected by checking the wave function symmetry at the band edges.

A typical example of the Dirac mass tuning is found in a modulated honeycomb lattice model, which involves spatial patterns in hopping amplitudes \cite{PhysRevLett.114.223901,Wu:2016aa,Kariyado2017}. The proposed pattern includes two values for the hoppings arranged in a manner preserving $C_{6v}$ symmetry of the system, and it has been shown that a band inversion between states with $p$-wave symmetry and $d$-wave symmetry is realized by changing the ratio of the two hopping values. This method stands solely on the spatial patterning, and therefore, has been used to realize topologically nontrivial states in a wide variety of systems, ranging from an electronic artificial lattice \cite{PhysRevLett.124.236404} to macroscopic photonic crystals~\cite{Yang2018,Barik2018,Li2018,Shao2020,Parappurath2020}. 
However, the realization in real materials is limited because it is very difficult to tune hopping amplitudes at the atomic level. 
While introducing spatial patterns of holes in graphene has been proposed~\cite{PhysRevB.98.195416}, it is still very challenging since it requires appropriate precursors in bottom-up methods or lithography techniques with atomic precision in top-down methods. 
Although graphyne and graphdiyne are predicted to show unbalanced parity indices at inversion-symmetric momenta~\cite{Liu2019a,Sheng2019a}, the edge morphologies that exhibit topological edge states do not match the precursors in experiments~\cite{Li2010,Matsuoka2017,Hu2022a}.
It might be more feasible to perform patterned hydrogenation~\cite{Elias2009,Balog2010,Haberer2011,Chen2018,Song2022}, where hydrogen atoms eliminate the $\pi$ electrons of carbon atoms by changing the hybridization from $sp^2$ to $sp^3$ and result in effective holes. 
Recently, hydrogen adatoms on graphene can be deposited, laterally moved, and removed with atomic precision by using the scanning tunneling microscopy (STM) tip under varying sample voltages~\cite{Gonzalez-Herrero2016}.

In this paper, we theoretically demonstrate a band inversion in graphene with designed spatial patterns of hydrogenation. With our designs, graphene acquires band gaps of order of 1~eV. We further find that by changing the alignment of the hydrogenation while keeping the period, the band inversion is realized. This is confirmed in the first-principles density-functional-theory (DFT) calculations. We successfully build a tight-binding (TB) model describing the hydrogenated graphene. Using the TB model, we demonstrate gapless topological edge states and interface states as protected by the mirror symmetry and sublattice symmetry. The edge states are simulated in a nanoribbon geometry, while the interface states are simulated in patchwork geometry in which two regions with different hydrogenation patterns are aligned side by side. 
We discuss that the helical edge/interface states can be considered as Majorana-like modes with particle-hole symmetry.

\begin{figure*}[t]
    \centering
    \includegraphics[width=0.7\textwidth]{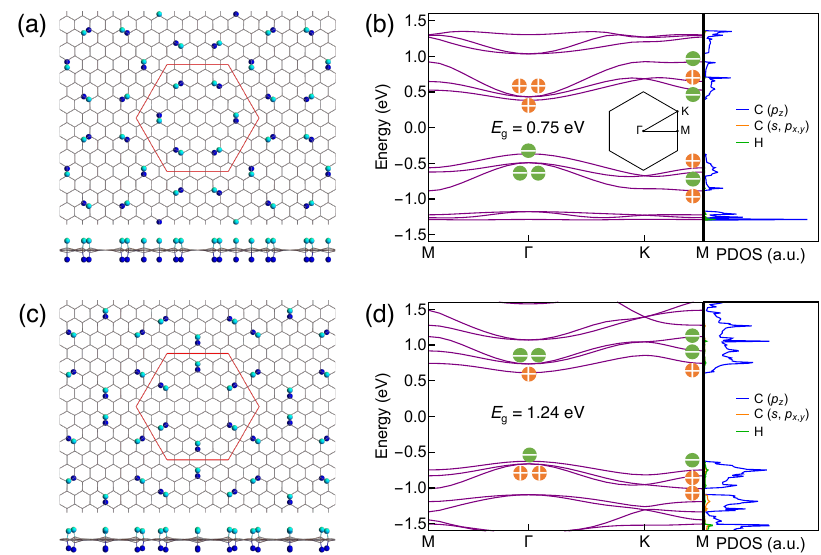}
    \caption{(a)~Stable structure of circular-patterned hydrogenated graphene obtained by DFT calculations, where only hydrogen atoms (cyan/blue for the ones above/below graphene) are shown for emphasizing their positions. The hexagonal unit cell is colored in red. (b)~DFT band structure of (a) with projected density of states (PDOS), where the parity of the eigenstates at $\Gamma$- and M-points for the six bands near the Fermi level is denoted by plus/minus sign. Inset: Brillouin zone. (c)~and (d) Same as (a) and (b) except for a radial pattern. Particle-hole symmetry is observed to good approximation for states around the band gap in both cases.}
    \label{fig:DFT}
\end{figure*}

\section{Results and Discussions}
\subsection{First-Principles Calculations}
Here, we conduct lattice relaxations and band structure calculations on hydrogenated graphene using the DFT method. 
We focus on the double-sided hydrogenation because it is energetically favorable as indicated by DFT calculations~\cite{Boukhvalov2008} and has been experimentally observed in both suspended graphene~\cite{Elias2009} and graphene on substrate with an annealing treatment~\cite{Chen2018,Song2022}.
Two hydrogenation patterns are considered as shown in Figs.~\ref{fig:DFT}(a) and \ref{fig:DFT}(c), where lattice structures are already relaxed. We name the patterns in Figs.~\ref{fig:DFT}(a) and \ref{fig:DFT}(c) circular pattern and radial pattern, respectively, for later use. Note that a similar design was used to realize topologically nontrivial states in a photonic crystal \cite{WangHu+2020+3451+3458}. The circular and radial patterns share the same superstructure period, namely, both of the patterns have 96 carbon atoms in a unit cell. Also, the numbers of adsorbed hydrogen atoms are the same for the two patterns.

Figures~\ref{fig:DFT}(b) and \ref{fig:DFT}(d) are the corresponding band structures with the projected density of states (PDOS). There are clear energy gaps for both of the circular (0.75 eV) and radial (1.24 eV) patterns. The obtained PDOS confirms that the states around the gap are mostly from $\pi$-electrons ($p_z$-orbitals), for both of the patterns. 
\begin{figure*}[t]
    \centering
    \includegraphics[width=0.85\textwidth]{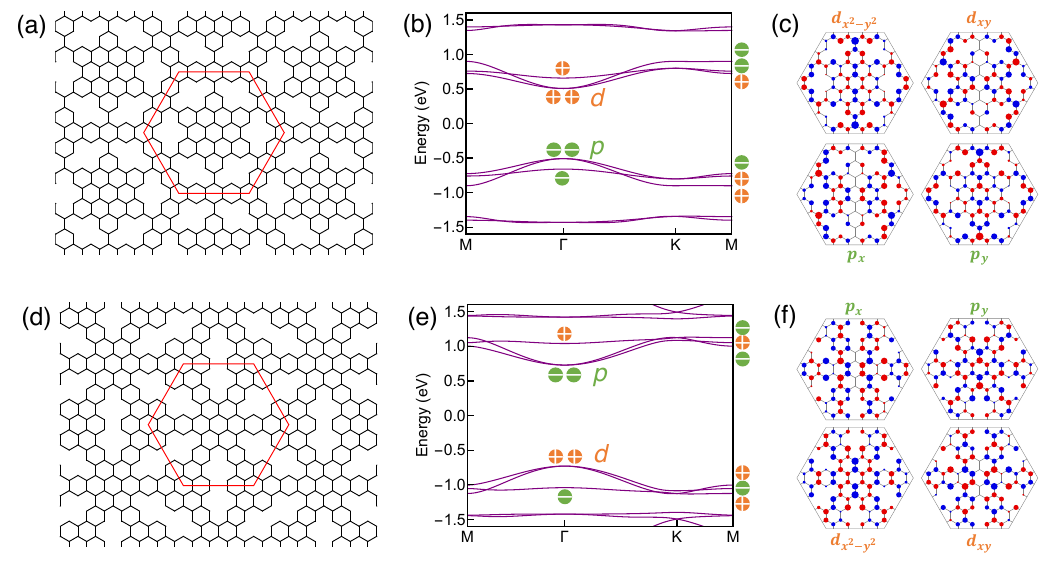}
    \caption{(a)~Schematic structure for graphene with a circular-patterned nanohole array, where the hexagonal unit cell is shown in red. (b)~Band structure of (a), where $p$ and $d$ states at the $\Gamma$-point are labeled explicitly, and the parity of the eigenstates at $\Gamma$- and M-points for the six bands near the Fermi level is denoted by plus/minus sign. (c)~Wave functions of $p$ and $d$ states labeled in (b), with amplitude and sign denoted by area and color of dots, respectively. (d)--(f) Same as (a)--(c) except for a radial pattern. Particle-hole symmetry appears in the TB model as guaranteed by the sublattice symmetry. It is noted that the naming of $p$ and $d$ states is based on the symmetry of the eigen wave function in the unit cell.}
    \label{fig:TB}
\end{figure*}

When it comes to the symmetry characters of the wave functions that are essential for detecting the band inversion, we observe a difference between the circular and the radial patterns. 
The parity index ($C_2$ eigenvalue with respect to the center of hexagonal unit cell) for all valence bands in the circular pattern is $(N_{\Gamma}^+, N_{\text{M}}^+)=(97,99)$, while the one in the radial pattern is $(N_{\Gamma}^+, N_{\text{M}}^+)=(99,99)$ (note that the total number of valence states is 198 in both cases).
The imbalance of parity index in the circular pattern indicates a nontrivial topology, while the radial pattern is trivial~\cite{Benalcazar2014,Noh2018}.
In Figs.~\ref{fig:DFT}(b) and \ref{fig:DFT}(d), the parity labels ($\pm$) are indicated for the three highest energy valence bands and the three lowest energy conduction bands at the $\Gamma$-point. 
For the circular pattern, the three valence bands are parity odd while the three conduction bands are parity even. In sharp contrast, for the radial pattern, the three valence bands contain two parity even states and one parity odd state, while the three conduction bands contain one parity even state and two parity odd states. The different order of the parity labels with respect to the band gap is indicative of the band inversion. To say more, the three highest valence bands in the circular pattern show a discrepancy in the distribution of parity labels between the $\Gamma$-point and the M-point, while those in the radial pattern do not. From this, one can conclude that the three highest valence bands fully capture the topology of the system.

\subsection{Tight-Binding Model}\label{sec:TB}
In order to have deeper understanding of the electronic properties of the hydrogenated graphene, and to reduce the computational difficulty for the edge/interface state studies, we construct a TB model. As the obtained PDOS in Figs.~\ref{fig:DFT}(b) and \ref{fig:DFT}(d) indicate, it is expected that the hydrogenated graphene can be modeled as a TB model where $\pi$-electrons are hopping around the network of the carbon atoms, just as the pristine graphene. 
However, we have to note that the hydrogenation gives a strong local potential, and effectively an electron cannot hop on the hydrogenated sites. 
Within the effective TB description, the hydrogenation simply eliminates the corresponding carbon site. 
To simplify the model, we only retain the nearest-neighbor hoppings, and assume that the hopping parameters are uniform, which gives our Hamiltonian
\begin{equation}
    H = -t\sum_{\langle i,j\rangle}c^\dagger_ic_j, \label{eq:TBhamilt}
\end{equation}
where the summation over the nearest-neighbor pair $\langle i,j\rangle$ is taken over the bonds written in Figs.~\ref{fig:TB}(a) and \ref{fig:TB}(d) for the circular pattern and the radial pattern, respectively. For the nearest-neighbor hopping, we set $t=2.7$ eV as in the pristine graphene~\cite{CastroNeto2007}.

Figures~\ref{fig:TB}(b) and \ref{fig:TB}(e) show the TB band structures for the circular and radial patterns, respectively. 
Despite the simplicity of the model, even the relative size of the band gaps is consistent with the DFT results shown in Figs.~\ref{fig:DFT}(a) and \ref{fig:DFT}(c), namely, the circular pattern (1.02~eV) yields a smaller gap than the radial pattern (1.46~eV). 
Importantly, the TB model successfully captures the topology of the system by showing the parity index for all valence bands $(N_{\Gamma}^+, N_{\text{M}}^+)=(19,21)$ in the circular pattern and $(N_{\Gamma}^+, N_{\text{M}}^+)=(21,21)$ in the radial pattern (note that the total number of valence states is 42 in both cases).
Moreover, the parity index for the three highest energy valence bands and three lowest energy conduction bands shown in Figs.~\ref{fig:TB}(b) and \ref{fig:TB}(e) successfully reproduce the DFT results shown in Figs.~\ref{fig:DFT}(b) and \ref{fig:DFT}(d).
It is worth noting that if we focus on the doubly degenerate bands (labeled by $p$ or $d$ in Fig.~\ref{fig:TB}), the similarity between the TB results and the DFT results is remarkable. For the topological characterization, doubly degenerate bands at the $\Gamma$-point where the band inversion occurs are essential, and therefore, the TB model can be used in the following arguments for the topological edge/interface states. For the nondegenerate states at the $\Gamma$-point (labeled by $+$ or $-$), there is some discrepancy in energy between the TB and the DFT results. This energy discrepancy does not affect topology, and will be resolved if we think of the spatial dependence of the hopping parameters and handle the hydrogenation more realistically, which is left for future study.

Now, we are in position to discuss the symmetry of the wave function. Again, we focus on the three highest energy valence bands and the three lowest energy conduction bands at the $\Gamma$-point. For the circular pattern, the doubly degenerate valence bands have $p$-like symmetry (more specifically, $p_x$- or $p_y$-like symmetry), while the doubly degenerate conduction bands have $d$-like symmetry (more specifically, $d_{x^2-y^2}$- or $d_{xy}$-like symmetry) [Fig.~\ref{fig:TB}(c)]. In contrast, for the radial pattern, the doubly degenerate valence bands have $d$-like symmetry, while the doubly degenerate conduction bands have $p$-like symmetry [Fig.~\ref{fig:TB}(f)]. This difference signals the band inversion.
As in the case of the DFT results, by comparing the distribution of the parity labels between the $\Gamma$-point and the M-point, it is concluded that the circular pattern is in the topologically nontrivial phase. 
Finally, we note that particle-hole symmetry in the band structures shown in Figs. \ref{fig:DFT}(b) and \ref{fig:DFT}(d) is well captured by the TB model with nearest-neighbor hopping where sublattice symmetry is guaranteed.

\subsection{Edge and Interface States}
\begin{figure*}[t]
    \centering
    \includegraphics[width=0.78\textwidth]{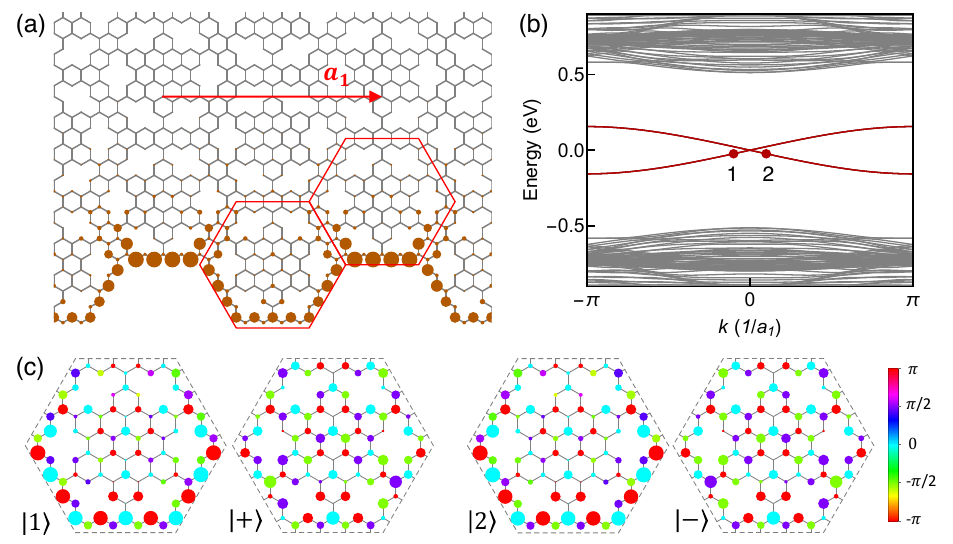}
    \caption{(a)~Geometry of graphene nanoribbon with a circular-patterned nanohole array and molecule-zigzag edge morphology for calculations of topological edge states. The local density of states $|\psi_i|^2$ for edge states 1 and 2 labeled in (b) are shown as brown dots. (b)~Energy dispersion of (a), with the edge states colored in red and states 1 and 2 at $k=\pm0.1\pi/a_1$ labeled explicitly. (c)~Comparison between the wave functions in the outmost unit cell of states 1 and 2 and the bulk pseudospin states $|\pm\rangle$ defined in equation~(\ref{eq:pseudospin}) in the text, with amplitude and phase denoted by area and color of dots, respectively. 
    }
    \label{fig:edge}
\end{figure*}

In order to see the topological nature of the obtained states, we rely on the idea of the bulk-edge correspondence, where topological nontriviality is reflected in the edge/interface spectrum. As shown in Fig.~\ref{fig:edge}, we use the TB model to investigate the band structure in a ribbon geometry, which is periodic and infinite in one direction while finite in the other direction. We set the infinite direction in parallel with the direction of the zigzag edge of the underlying graphene. 
In the finite direction, the open boundary condition is adopted, i.e., the hopping network of the TB model is truncated at a given boundary. 
In the above, the circular pattern has been identified as topologically nontrivial based on the distribution of the $C_2$ parity labels with respect to the center of the hexagonal unit cell shown in Fig.~\ref{fig:TB}(a), and the edge morphology is called molecule-zigzag~\cite{Kariyado2017} where the hexagonal unit cells are not broken, as displayed in Fig.~\ref{fig:edge}(a). 

Figure~\ref{fig:edge}(b) shows the band structure of the ribbon system as a function of the momentum along the ribbon for the circular pattern. Within the bulk band gap we clearly see two bands which cross each other with exactly zero energy gap at the $\Gamma$-point, indicative of the topological nontriviality of the state. Figure~\ref{fig:edge}(a) illustrates a typical local density of states for the edge state, confirming that the in-gap states are localized to the edge. The two crossing bands indicate that each of the branches corresponds to a right or left moving state, just like helical edge states in the quantum spin-Hall (QSH) effect where opposite spins flow in opposite directions. In our case, while the Hamiltonian~(\ref{eq:TBhamilt}) does not have spin degrees of freedom, the pseudospin defined using the wave function patterns takes over the roles of the real spin in the QSH state. 
In the present system, the zero energy gap in the helical edge states is guaranteed by a nonzero mirror winding number protected by sublattice symmetry and mirror symmetry simultaneously~\cite{Kariyado2017}.
Characterized by particle-hole symmetry, the topological edge/interface states in Fig.~\ref{fig:edge}(b) may be considered as Majorana-like modes.
For the armchair edge (not discussed explicitly here), as the mirror operation exchanges the A-B sublattices, a minigap will be opened at $k=0$.

As we have noted, $p_x$-, $p_y$-, $d_{x^2-y^2}$-, and $d_{xy}$-like bands are important for describing the bulk band structure near zero energy. With the double degeneracy, the pseudospin up ($+$) and down  ($-$) states in the bulk are defined as
\begin{align}
    |p_{\pm}\rangle &= (|p_x\rangle \pm i |p_y\rangle)/\sqrt{2}, \\
    |d_{\pm}\rangle &= (|d_{x^2-y^2}\rangle \pm i |d_{xy}\rangle)/\sqrt{2},
\end{align}
for $p$- and $d$-like bands, respectively. In order to make comparison with the edge states, it is convenient to define bulk pseudospin states $|\pm\rangle$ as \cite{PhysRevB.98.195416}
\begin{equation}\label{eq:pseudospin}
    |\pm\rangle = (|p_{\pm}\rangle \pm i |d_{\pm}\rangle)/\sqrt{2}.
\end{equation}
Figure~\ref{fig:edge}(c) compares $|\pm\rangle$ with the edge wave functions for typical right-moving ($|1\rangle$) or left-moving ($|2\rangle$) states. Although there is discrepancy in amplitude (dot areas in the figure), the phase distribution (colors in the figure) is in very good agreement between the bulk and edge states, which means that the edge states carry the pseudospins, and the right-moving and the left-moving states have the opposite pseudospins. 

\begin{figure*}[t]
    \centering
    \includegraphics[width=0.78\textwidth]{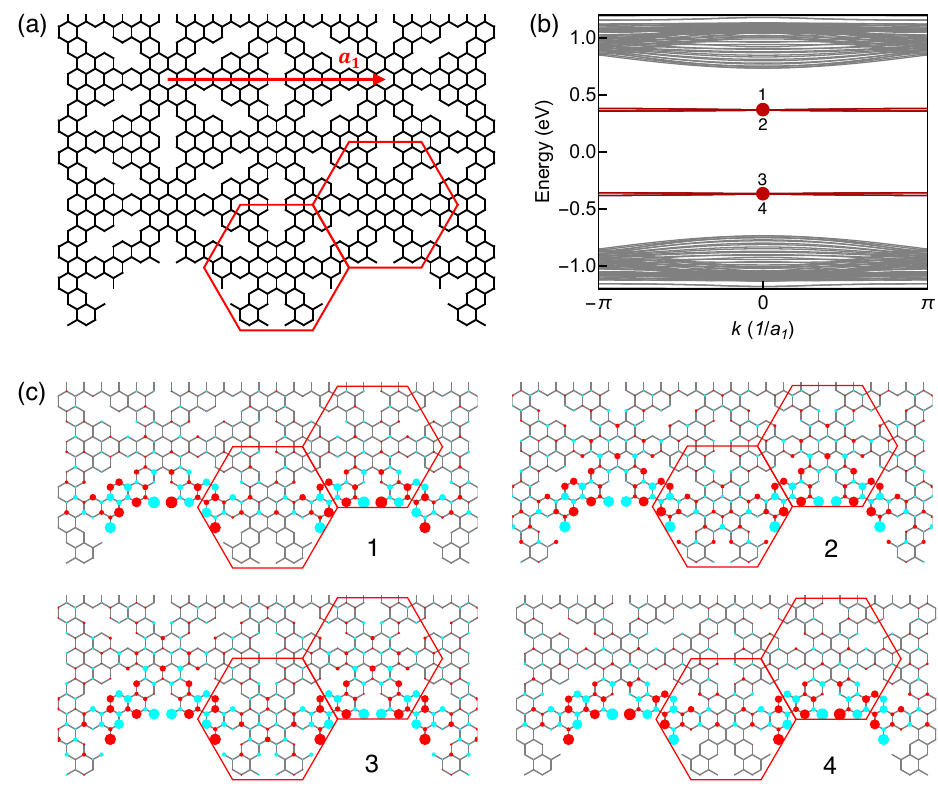}
    \caption{(a)~Geometry of graphene nanoribbon with a radial-patterned nanohole array and molecule-zigzag edge morphology for calculations of edge states. (b)~Energy dispersion of (a), with the edge states colored in red and states 1 to 4 at $k=0$ labeled from high energy to low energy. (c)~Wave functions of states 1 to 4, with amplitude and sign denoted by area and color of dots, respectively.
    }
    \label{fig:edge_radial}
\end{figure*}

We also investigate the ribbon system of the radial pattern shown in Fig.~\ref{fig:edge_radial}(a), which is topologically trivial as discussed in Sec.~\ref{sec:TB}. 
As can be seen in Fig.~\ref{fig:edge_radial}(b), the energy dispersion of the ribbon shows four nearly dispersionless states with finite energy inside the bulk gap. 
In Fig.~\ref{fig:edge_radial}(c) we display the wave functions of these four states at the $\Gamma$-point, which show a good localization at the edge. 
These edge states mainly localize in the concave regions and the overlap between two neighboring concave regions is small, giving the flatness of the energy dispersion.
The nearly dispersionless edge states are in sharp contrast to the helical edge states in the ribbon of the circular pattern shown in Figs.~\ref{fig:edge}(b) and \ref{fig:edge}(c), which originate from the nontrivial topology.

\begin{figure*}[t]
    \centering
    \includegraphics[width=0.7\textwidth]{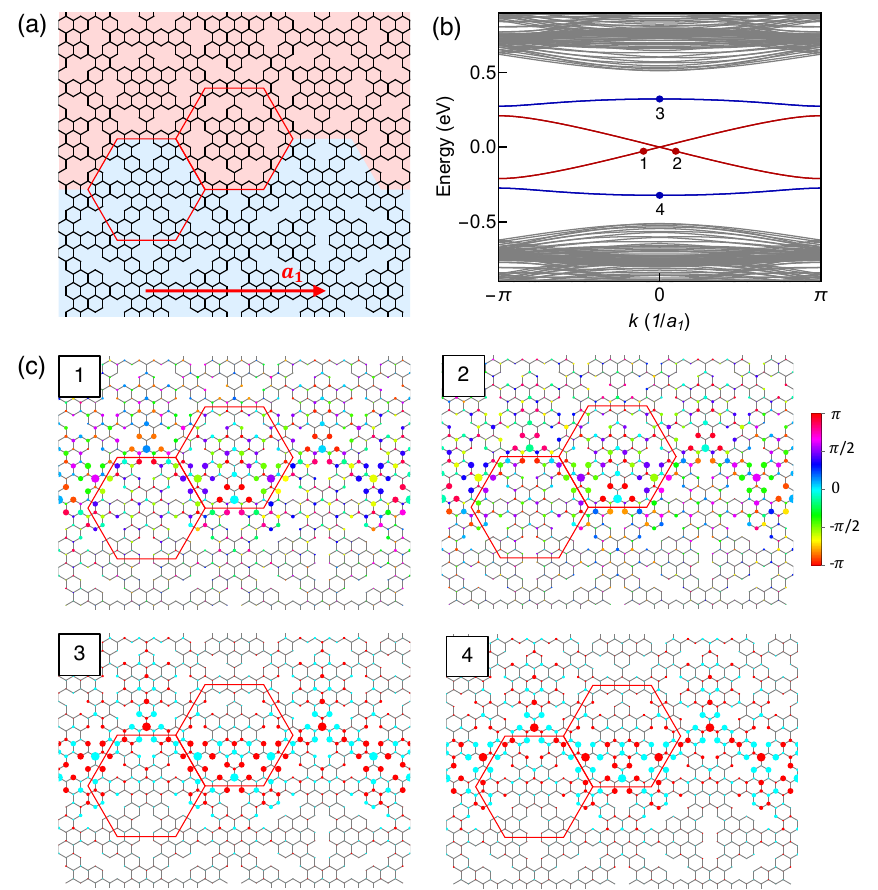}
    \caption{(a)~Graphene patchwork with a molecule-zigzag interface between circular- (red) and radial-patterned (blue) nanohole arrays. (b)~The energy dispersion of (a) with helical and nearly dispersionless edge states colored in red and blue, respectively. States 1 and 2 at $k=\pm0.1\pi /a_1$ and states 3 and 4 at $k=0$ are labeled explicitly. (c)~Wave functions of states 1 to 4, with amplitude and sign denoted by area and color of dots, respectively.}
    \label{fig:interface}
\end{figure*}

Interface states can also be used to characterize topology of a system through the bulk-interface correspondence. For this purpose, we consider a patchwork structure consisting of two regions with the circular and radial patterns. The patchwork is stripelike in the sense that each of the two regions occupies a finite width quasi one-dimensional (1D) region, and the entire system is regarded as a stack of the quasi-1D regions. Figure~\ref{fig:interface}(a) illustrates the patchwork structure near the interface of the two regions. Same as in the case of the edge, namely the hydrogenated graphene exposed to vacuum as shown in Figs.~\ref{fig:edge}(a) and \ref{fig:edge_radial}(a), the zigzag direction of the underlying graphene is chosen as the direction of the interface. Then, the TB model is used to calculate the band structure as a function of the momentum in parallel with the interface [horizontal direction in Fig.~\ref{fig:interface}(a)] and the result is shown in Fig.~\ref{fig:interface}(b). Again, within the bulk energy gap helical interface states are observed [red lines in Fig.~\ref{fig:interface}(b)], confirming that the circular and radial patterns are topologically distinct from each other. 
In addition, there also arises weakly dispersive bands in the gap [blue lines in Fig.~\ref{fig:interface}(b)].  
In Fig.~\ref{fig:interface}(c) we display the wave functions for the four in-gap states. The helical interface states 1 and 2 originate from the nontrivial topology in the circular pattern; while the nearly dispersionless states 3 and 4 inherit from the mirror-even edge states in the radial pattern, as indicated by the similarity between the wave functions of the bottom-left unit cell in states 3 and 4 in Fig.~\ref{fig:interface}(c) and the wave functions of the upper-right unit cell in states 2 and 3 in Fig.~\ref{fig:edge_radial}(c).
Transport properties of the system are dominated by the helical interface states when the chemical potential is well tuned where the influences from the nearly dispersionless states are expected to be small.

\section{Summary and Outlook}
To summarize, we have investigated band structures of graphene with designed hydrogenation by means of the first-principles calculations and revealed that appropriate designs lead to topologically distinct states in hydrogenated graphene. The topological nontriviality is detected by the imbalance of the $C_2$ parity indices. It has also been confirmed that a simple tight-binding model with the nearest-neighbor hoppings where the hydrogenated carbon sites are modeled as eliminated sites is sufficient to capture the topological character of the system. Using the tight-binding model, the bulk-edge and bulk-interface correspondence is demonstrated by calculating the band structure using the ribbon geometry and the stripelike geometry, respectively. In both of the cases, within the bulk gap we have found a pair of helical states with exactly zero energy gap, signaling the topological nontriviality of the system. The edge/interface states carrying opposite pseudospins propagate in opposite directions, which indicates that the hydrogenated graphene is a promising candidate for topological edge/interface state based devices. 
Exhibiting particle-hole symmetry, these helical edge/interface states can be considered as Majorana-like modes.
In addition to hydrogenation, fluorination of graphene~\cite{Robinson2010,Nair2010,Zboril2010,Lee2012} is expected to derive similar results.
The technique of manipulating hydrogen atoms on graphene with atomic precision~\cite{Gonzalez-Herrero2016} potentially can be utilized to rewrite the electronic topological circuits.

\section*{Methods}
First-principles calculations are performed within the DFT scheme using Vienna Ab-initio Simulation Package~\cite{Kresse1996}, where the projector augmented-wave method~\cite{Kresse1999} is implemented. The generalized gradient approximation of Perdew-Burke-Ernzerhof type~\cite{Perdew1996} is used to treat the exchange-correlation potential. The plane-wave cutoff energy is set to be 520~eV. The Brillouin zone is sampled with a $9\times9\times1$ $\Gamma$-centered mesh. Structural relaxations and electronic structure calculations are performed until the Hellmann-Feynman forces are smaller than $10^{-4}$~eV/{\AA} and the energy tolerances are below $10^{-6}$~eV/atom. A vacuum layer of 2~nm between graphene sheets is added so that the interlayer couplings are negligible. The post-processing is done using VASPKIT~\cite{Wang2021}, with parity indices obtained using VASP2Trace~\cite{Vergniory2019}.

\section*{Data availability}
All the data that support the findings of this study are available from the corresponding author upon reasonable request, following the policy of JST.

\bigskip
\section*{Code availability}
All the codes that support the findings of this study are available from the corresponding author upon reasonable request, following the policy of JST.

\bibliography{hydrogenated.bib}

\section*{Acknowledgment}
This work is supported by CREST, JST (Core Research for Evolutionary Science and Technology, Japan Science and Technology Agency) (Grant Number JPMJCR18T4).

\section*{Author contributions} 
X.H. conceived and supervised the project. Y.C.J. performed the calculations. All authors joined to write the manuscript and fully contributed to the project.

\section*{Declaration of interests} 
The authors declare no competing financial and non-financial interests.

\end{document}